\documentclass[aps,twocolumn,floatfix,showpacs]{revtex4}
\usepackage{graphicx}
\usepackage{amsmath}
\usepackage{amssymb}
\usepackage{bm}

\begin{document}

\title{Quantum point contact as a probe of a topological superconductor}
\author{M. Wimmer}
\affiliation{Instituut-Lorentz, Universiteit Leiden, P.O. Box 9506, 2300 RA Leiden, The Netherlands}
\author{A. R. Akhmerov}
\affiliation{Instituut-Lorentz, Universiteit Leiden, P.O. Box 9506, 2300 RA Leiden, The Netherlands}
\author{J. P. Dahlhaus}
\affiliation{Instituut-Lorentz, Universiteit Leiden, P.O. Box 9506, 2300 RA Leiden, The Netherlands}
\author{C. W. J. Beenakker}
\affiliation{Instituut-Lorentz, Universiteit Leiden, P.O. Box 9506, 2300 RA Leiden, The Netherlands}

\date{January 2011}
\begin{abstract}
We calculate the conductance of a ballistic point contact to a superconducting wire, produced by the \textit{s}-wave proximity effect in a semiconductor with spin-orbit coupling in a parallel magnetic field. The conductance $G$ as a function of contact width or Fermi energy shows plateaus at half-integer multiples of $4e^{2}/h$ if the superconductor is in a topologically nontrivial phase. In contrast, the plateaus are at the usual integer multiples in the topologically trivial phase. Disorder destroys all plateaus except the first, which remains precisely quantized, consistent with previous results for a tunnel contact. The advantage of a ballistic contact over a tunnel contact as a probe of the topological phase is the strongly reduced sensitivity to finite voltage or temperature. 
\end{abstract}
\pacs{73.23.Ad, 73.23.-b, 74.25.fc, 74.45.+c}
\maketitle

\section{Introduction}
\label{intro}

Massless Dirac fermions have the special property that they can be confined without the energy cost from zero-point motion. In graphene, this manifests itself as a Landau level at zero energy, without the usual $\frac{1}{2}\hbar\omega_{c}$ offset \cite{McC56}. The zeroth Landau level contributes half as much to the Hall conductance as the higher levels (because it is already half-filled in equilibrium), leading to the celebrated half-integer quantum Hall plateaus \cite{Nov05,Zha05}. In a semiclassical description, the $\pi$ phase shift at turning points, responsible for the zero-point energy, is canceled by the Berry phase of $\pi$, characteristic for the periodic orbit of a Dirac fermion.

The same absence of zero-point energy appears when Dirac fermions are confined by superconducting barriers, produced by the proximity effect in a topological insulator \cite{Jac81,Fu08}. Because of particle-hole symmetry in a superconductor, a state at zero excitation energy is a Majorana bound state, with identical creation and annihilation operators. A superconductor that supports Majorana bound states is called topological \cite{Has10,Qi10}.

Tunneling spectroscopy is a direct method of detection of a topological superconductor \cite{Bol07,Law09,Fle10,Fle11}. Resonant tunneling into a Majorana bound state produces a conductance of $2e^{2}/h$, while without this state the tunneling conductance vanishes \cite{Law09}. The tunneling resonance becomes broader if the tunneling probability is increased, and one might surmise that the resonance would vanish if the conductance is measured via a ballistic contact. We show in this paper, by means of a model calculation, that the contrary is true: The signature of the topological phase is more robust when measured by a ballistic contact than by a tunnel contact.

Our model calculation is in accord with general theoretical considerations \cite{Ber09a,Ber09b}, but may appear counter-intuitive. After all, the Majorana bound state no longer exists as an individual energy level if it is connected by a ballistic contact to a normal metal, since the level broadening then exceeds the level spacing. As we have found, the topological phase of the superconductor still manifests itself in the conductance of a ballistic point contact, in a way reminiscent of the half-integer quantum Hall plateaus.

\begin{figure}[tb]
\centerline{\includegraphics[width=1\linewidth]{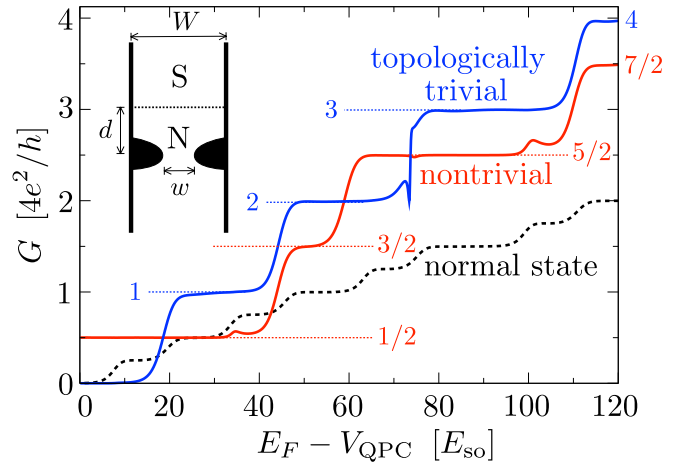}}
\caption{\label{fig_QPC1}
Solid curves: conductance of a ballistic normal-metal--superconductor (NS) junction, with the superconductor in a topologically trivial phase (blue curve, $\Delta=8\,E_{\rm so}$) or nontrivial phase (red curve, $\Delta=4\,E_{\rm so}$). The black dashed curve is for an entirely normal system ($\Delta=0$). The data is obtained from the model Hamiltonian \cite{Lut10a,Ore10} of a semiconducting wire on a superconducting substrate in a parallel magnetic field (Zeeman energy $E_{Z}=6\,E_{\rm so}$), for the ballistic point contact geometry shown in the inset (not to scale, $d=2.5\,l_{\rm so}$, $W=l_{\rm so}$). By varying the potential $V_{\rm QPC}$ at constant Fermi energy $E_{F}=120\,E_{\rm so}$, the point contact width $w$ is varied between $0$ and $W$. The dotted horizontal lines indicate the shift from integer to half-integer conductance plateaus upon transition from the topologically trivial to nontrivial phase.
}
\end{figure}

\section{Integer versus half-integer conductance plateaus}
\label{halfintegerinteger}

We consider the model Hamiltonian \cite{Lut10a,Ore10} of a two-dimensional semiconducting wire with an \textit{s}-wave proximity-induced superconducting gap $\Delta$. (See App.\ \ref{app_model} for a detailed description.) We have calculated the scattering matrix of a quantum point contact (QPC) in the normal region (N) at a distance $d$ from the superconducting region (S), by discretizing the Hamiltonian on a square lattice (lattice constant $a=l_{\rm so}/40$, with $l_{\rm so}$ the spin-orbit scattering length). Our key result is presented in Fig.\ \ref{fig_QPC1}. The number of propagating modes in the point contact (and hence the transmittance $T_{\rm QPC}$) is varied by changing the electrostatic potential $V_{\rm QPC}$ inside the point contact, at constant Fermi energy $E_{F}$. Spin degeneracy is removed by the Zeeman energy $E_{Z}=\frac{1}{2}g\mu_{B}B$ in a magnetic field $B$ (parallel to the wire), so that when the entire system is in the normal state ($\Delta\rightarrow 0$) the conductance increases step wise in units of $e^{2}/h$ (black dashed curve, showing the step wise increase of the transmittance from $T_{\rm QPC}=0$, for a fully pinched off contact, to $T_{\rm QPC}=8$, for a maximally open contact). 

The conductance $G$ of the NS junction is obtained from the Andreev reflection eigenvalues $R_{n}$ at the Fermi level,
\begin{equation}
G=\frac{2e^{2}}{h}\sum_{n}R_{n}(E_{F}).\label{BTK}
\end{equation}
The factor of two accounts for the fact that charge is added to the superconductor as Cooper pairs of charge $2e$. (The spin degree of freedom is included in the sum over $n$.) The superconductor can be in a topologically trivial ($Q=1$) or nontrivial ($Q=-1$) phase, depending on the relative magnitude of $E_{Z}$, $\Delta$, and the spin-orbit coupling energy $E_{\rm so}=\hbar^{2}/m_{\rm eff}l_{\rm so}^{2}$. The blue and red solid curves show these two cases, where the topological quantum number $Q={\rm sign}\,{\rm Det}\,r$ was obtained in an independent calculation from the determinant of the reflection matrix \cite{Mer02,Akh11,Ful11}. As we see from Fig.\ \ref{fig_QPC1}, the conductance shows plateaus at values $G_{p}$, $p=0,1,2,\ldots$, given by
\begin{equation}
G_{p}=\frac{4e^{2}}{h}\times\left\{\begin{array}{ll}
p&{\rm if}\;\;Q=1,\\
p+1/2&{\rm if}\;\;Q=-1.
\end{array}\right.\label{Gpdef}
\end{equation}

The sequence of conductance plateaus in the topologically trivial and nontrivial phases can be understood from basic symmetry requirements. As discovered by B\'{e}ri \cite{Ber09b}, particle-hole symmetry requires that the $R_{n}$'s at the Fermi level are either twofold degenerate or equal to 0 or 1. (See App.\ \ref{app_Beri} for a derivation.) A nondegenerate unit Andreev reflection eigenvalue is therefore pinned to exactly this value and contributes to the conductance a quantized amount of $2e^{2}/h$. This is the signature of the topological superconductor which persists even after the Majorana bound state has merged with the continuum of states in the normal metal contact. 

If we include only the degenerate $R_{n}$'s in the sum over $n$ (indicated by a prime, ${\sum}'$), we may write
\begin{equation}
G=\frac{e^{2}}{h}\left(1-Q+4{\sum}'_{n}R_{n}\right).\label{GQ}
\end{equation}
A new mode that is fully Andreev reflected thus adds $4e^{2}/h$ to the conductance, with an offset of $0$ or $2e^{2}/h$ in the topologically trivial or nontrivial phases. The resulting conductance plateaus therefore appear at integer or half-integer multiples of $4e^{2}/h$, depending on the topological quantum number, as expressed by Eq.\ \eqref{Gpdef} and observed in the model calculation.

The quantum point contact conductance plateaus in the topologically nontrivial phase occur at the same half-integer multiples of $4e^{2}/h$ as the quantum Hall plateaus in graphene, but the multiplicity of $4$ has an entirely different origin: In graphene, the factor of four accounts for the twofold spin and valley degeneracy of the energy levels, while in the NS junction there is no degeneracy of the energy levels. One factor of two accounts for the Cooper pair charge, while the other factor of two is due to the B\'{e}ri degeneracy of the non-unit Andreev reflection eigenvalues.

\section{Effect of disorder}
\label{disorder}

\begin{figure}[tb]
\centerline{\includegraphics[width=0.9\linewidth]{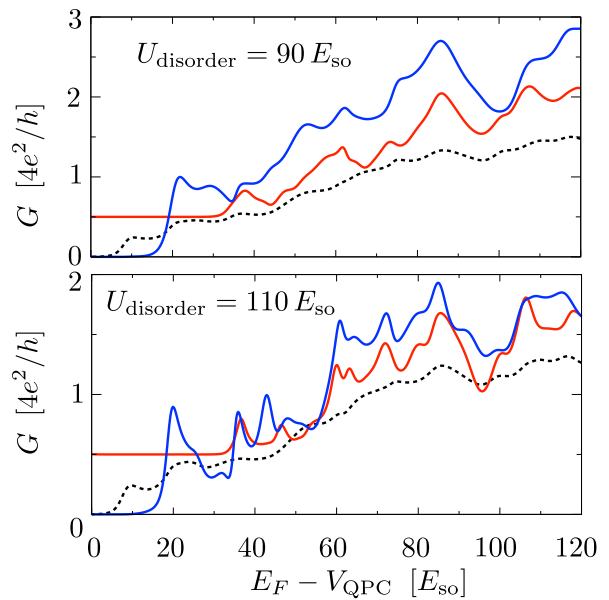}}
\caption{\label{fig_QPC2}
Same as Fig.\ \ref{fig_QPC1}, but now in the presence of disorder (for two values of the disorder strength). The first conductance plateau in the topologically nontrivial phase remains precisely quantized. 
}
\end{figure}

While in the quantum Hall effect all plateaus are insensitive to disorder, in the NS junction this applies only to the first plateau. As follows from Eq.\ \eqref{GQ}, the first plateau at $G=(1-Q)(e^{2}/h)$ is determined by the topological quantum number $Q$, which is robust against perturbations of the Hamiltonian. No such topological protection applies to the higher plateaus, since $R_{n}$ can take on any value between $0$ and $1$ in the presence of disorder. 

This is demonstrated in Fig.\ \ref{fig_QPC2}, where we have added disorder to the model calculation (both in the normal and in the superconducting region), by randomly chosing the electrostatic potential at each lattice point from the interval $[-U_{\rm disorder},U_{\rm disorder}]$. The mean free path $l_{\rm mfp}\propto U_{\rm disorder}^{-2}$ depends rather sensitively on the disorder strength. We show results for $U_{\rm disorder}=90\,E_{\rm so}$ and $110\,E_{\rm so}$, when the mean free path (calculated in Born approximation) is estimated at $l_{\rm mfp}=9\,l_{\rm so}$ and $6\,l_{\rm so}$, respectively. (The topologically nontrivial phase itself persists up to $l_{\rm mfp}= 3\,l_{\rm so}$.)

\section{Effect of finite voltage and temperature}
\label{sec_voltage}

These are all results in the limit of zero applied voltage $V$ and zero temperature $T$. There is then no qualitative difference between the $2e^{2}/h$ conductance resonance in the tunneling regime or in the ballistic regime. A substantial difference appears at finite voltages or temperatures.

\begin{figure}[tb]
\centerline{\includegraphics[width=0.9\linewidth]{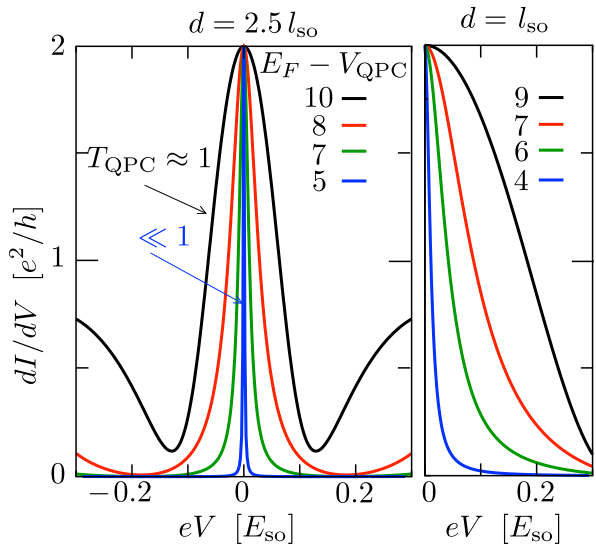}}
\caption{\label{fig_QPC3}
Differential conductance at different values of $E_{F}-V_{\rm QPC}$ (listed in units of $E_{\rm so}$), for two values of the distance $d$ between quantum point contact and superconductor. The data is taken on the first conductance plateau in the topologically nontrivial phase ($\Delta=4\,E_{\rm so}$, $U_{\rm disorder}=90\,E_{\rm so}$). The quantum point contact is in the tunneling regime for the blue curve (transmittance $T_{\rm QPC}=0.1$) and in the single-mode ballistic regime for the black curve ($T_{\rm QPC}\approx 1$). The width of the conductance peak increases both upon increasing $T_{\rm QPC}$ and upon decreasing $d$. 
}
\end{figure}

Considering first the effect of a nonzero applied voltage, we show in Fig.\ \ref{fig_QPC3} the differential conductance 
\begin{equation}
\frac{dI}{dV}=\frac{2e^{2}}{h}\sum_{n}R_{n}(E_{F}+eV).\label{BTKfiniteV}
\end{equation}
The peak centered at $V=0$ is the signature of the topologically nontrivial phase \cite{Law09}. The height $2e^{2}/h$ of this peak remains the same as $T_{\rm QPC}$ is raised from $0$ to $1$ by opening up the point contact, but the peak width increases. For a given $T_{\rm QPC}$, moving the point contact closer to the superconductor also has the effect of increasing the peak width (right panel in Fig.\ \ref{fig_QPC3}).

\begin{figure}[tb]
\centerline{\includegraphics[width=0.9\linewidth]{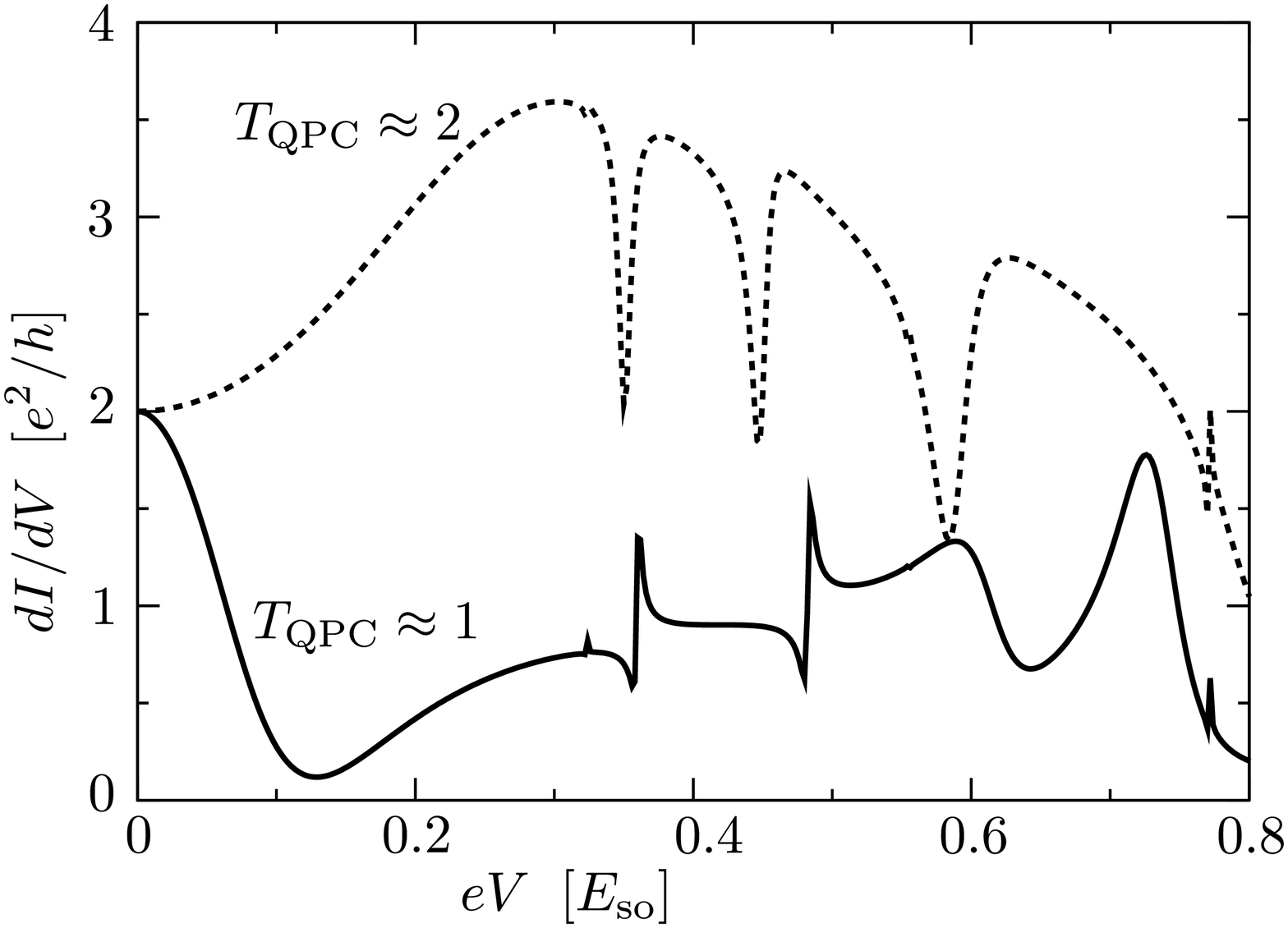}}
\caption{\label{fig_QPC4}
The solid curve is the same data as the black curve in the left panel of Fig.\ \ref{fig_QPC3} ($E_{F}-V_{\rm QPC}=10\,E_{\rm so}$, $T_{\rm QPC}\approx 1$), but on a larger voltage scale to show the resonances beyond the conductance peak centered at $V=0$. (The curve is $\pm V$ symmetric.) The dashed curve shows that the conductance peak becomes a conductance dip when a second mode opens up in the quantum point contact ($E_{F}-V_{\rm QPC}=20\,E_{\rm so}$, $T_{\rm QPC}\approx 2$).
}
\end{figure}

These considerations apply to the transition from the tunneling regime ($T_{\rm QPC}\ll 1$) to the ballistic regime with a single transmitted mode ($T_{\rm QPC}\approx 1$). If we open the point contact further, a second mode is partially transmitted and at $T_{\rm QPC}\approx 1.3$ the conductance peak switches to a conductance dip. Fig.\ \ref{fig_QPC4} contrasts the inverted resonances at $T_{\rm QCP}$ equal to $1$ (conductance peak) and equal to $2$ (conductance dip). The voltage scale in this figure is larger than Fig.\ \ref{fig_QPC3}, to show also the higher-lying resonances.

A simple estimate for the width $\delta\simeq \hbar/\tau_{\rm dwell}$ of the conductance peak in the tunneling regime equates it to the inverse of the dwell time $\tau_{\rm dwell}$ of an electron (effective mass $m_{\rm eff}$) in the region (of size $W\times d$) between the point contact and the NS interface. For the relatively large mean free paths in the calculation ($l_{\rm mfp}>W,d$), the dwell time for point contact widths $w\ll W,d$ is given by $\tau_{\rm dwell}\simeq m_{\rm eff}Wd/\hbar T_{\rm QPC}$, so we estimate
\begin{equation}
\delta\simeq \frac{\hbar^{2} T_{\rm QPC}}{m_{\rm eff}Wd}=\frac{l_{\rm so}^{2}}{Wd}\,T_{\rm QPC}E_{\rm so}.\label{deltaestimate}
\end{equation}
This formula (without numerical prefactors) qualitatively accounts for the increase of $\delta$ with decreasing $d$ and with increasing $T_{\rm QPC}$ in the tunneling regime $T_{\rm QPC}\ll 1$, but for a quantitative description of the ballistic regime, including the switch from peak to dip, a more complete theory is needed. 

\begin{figure}[tb]
\centerline{\includegraphics[width=0.9\linewidth]{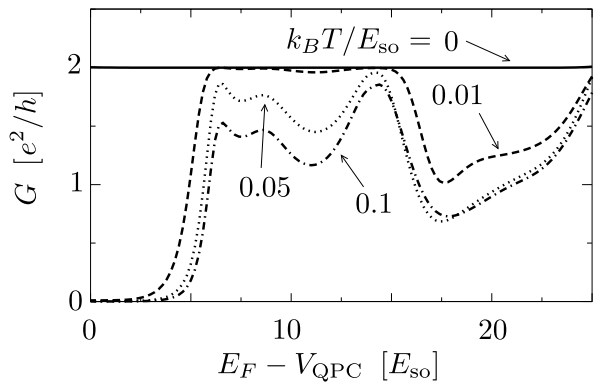}}
\caption{\label{fig_QPC5}
Conductance in the topologically nontrivial phase for different values of the thermal energy $k_{B}T$. The $2e^{2}/h$ plateau is suppressed at the smallest temperatures in the tunneling regime, and only for larger temperatures in the ballistic regime. 
}
\end{figure}

A similarly different robustness in the tunneling and ballistic regime appears if we consider the effect of a nonzero thermal energy $k_{B}T$ on the $2e^{2}/h$ conductance plateau. The finite-temperature conductance is calculated from
\begin{equation}
G(k_{B}T)=\frac{2e^{2}}{h}\int_{-\infty}^{\infty}dE\,\sum_{n}R_{n}(E)\frac{d}{dE}\frac{-1}{1+e^{E/k_{B}T}}.\label{BTKfiniteT}
\end{equation}
We show in Fig.\ \ref{fig_QPC5} how raising the temperature suppresses the $2e^{2}/h$ conductance plateau in the topologically nontrivial phase. The characteristic temperature scale for the suppression is $k_{B}T\simeq\delta$, so the plateau persists longest for $T_{\rm QPC}\approx 1$, when the line width $\delta$ of the resonance is the largest. 

\section{Conclusion}
\label{conclude}

In conclusion, we have presented a model calculation that shows how a quantum point contact can be used to distinguish the topologically trivial and nontrivial phases of a superconducting wire. The $2e^{2}/h$ conductance resonance in the tunneling regime \cite{Law09} persists in the ballistic regime, with a greatly reduced sensitivity to finite voltages and temperatures. The characteristic temperature scale (for a typical value $E_{\rm so}=0.1\,{\rm meV}$ of the spin-orbit coupling energy in InAs) reaches the 100\,mK range in the ballistic regime, which is still quite small but within experimental reach.

As more and more modes are opened in the ballistic point contact, new conductance plateaus appear at multiples of $4e^{2}/h$ which are integer in the trivial and half-integer in the nontrivial phase. This sequence of plateaus is a striking demonstration of the role which the degeneracy of Andreev reflection eigenvalues plays in the classification of topological superconductors \cite{Ber09b,Bee10}.

\acknowledgments

This research was supported by the Dutch Science Foundation NWO/FOM, by the Deutscher Akademischer Austausch Dienst DAAD, and by an ERC Advanced Investigator Grant.

\appendix

\section{Model Hamiltonian}
\label{app_model}

Our model calculations are based on the Hamiltonian of Refs.\ \cite{Lut10a,Ore10}, which describes an InAs nanowire on an Al or Nb substrate. The Bogoliubov-De Gennes Hamiltonian
\begin{align}
{\cal H}&=\begin{pmatrix}
1&0\\
0&\sigma_{y}
\end{pmatrix}
\begin{pmatrix}
H_{\rm R}-E_{F}&\Delta\\
\Delta^{\ast}&E_{F}-\sigma_{y}H_{\rm R}^{\ast}\sigma_{y}
\end{pmatrix}
\begin{pmatrix}
1&0\\
0&\sigma_{y}
\end{pmatrix}\nonumber\\
&=\begin{pmatrix}
H_{\rm R}-E_{F}&\Delta\sigma_{y}\\
\Delta^{\ast}\sigma_{y}&E_{F}-H_{\rm R}^{\ast}
\end{pmatrix}
\label{HBdG}
\end{align}
couples electron and hole excitations near the Fermi energy $E_{F}$ through an \textit{s}-wave superconducting order parameter $\Delta$. (We have made a unitary transformation to ensure that the condition for particle-hole symmetry has the form used in App.\ \ref{app_Beri}.)

The excitations are confined to a wire of width $W$ in the $x-y$ plane of the semiconductor surface inversion layer, where their dynamics is governed by the Rashba Hamiltonian
\begin{equation}
H_{\rm R}=\frac{\bm{p}^{2}}{2m_{\rm eff}}+U(\bm{r})+\frac{\alpha_{\rm so}}{\hbar}(\sigma_{x}p_{y}-\sigma_{y}p_{x})+\tfrac{1}{2}g_{\rm eff}\mu_{B}B\sigma_{x}.\label{HRashba}
\end{equation}
The spin is coupled to the momentum $\bm{p}=-i\hbar\partial/\partial{\bm r}$ by the Rashba effect, and polarized through the Zeeman effect by a magnetic field $B$ parallel to the wire (in the $x$-direction). Characteristic length and energy scales are $l_{\rm so}=\hbar^{2}/m_{\rm eff}\alpha_{\rm so}$ and $E_{\rm so}=m_{\rm eff}\alpha_{\rm so}^{2}/\hbar^{2}$. Typical values in InAs are $l_{\rm so}=100\,{\rm nm}$, $E_{\rm so}=0.1\,{\rm meV}$, $E_{Z}=\frac{1}{2}g_{\rm eff}\mu_{B}B=1\,{\rm meV}$ at $B=1\,{\rm T}$.

The electrostatic potential $U=U_{\rm QPC}+\delta U$ is the sum of a gate potential $U_{\rm QPC}$ and an impurity potential $\delta U$. The impurity potential $\delta U(x,y)$ varies randomly from site to site on a square lattice (lattice constant $a$), distributed uniformly in the interval $[-U_{\rm disorder},U_{\rm disorder}]$.

\begin{figure}[tb]
\centerline{\includegraphics[width=0.9\linewidth]{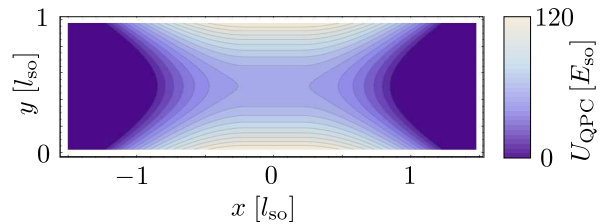}}
\caption{\label{fig_contour}
Contour plot of the quantum point contact potential \eqref{UQPC}, for the parameters $\ell=0.2\,l_{\rm so}$, $\hbar\omega_{x}=15\,E_{\rm so}$, $\hbar\omega_{y}=25\,E_{\rm so}$, $V_{\rm QPC}=55\,E_{\rm so}$. This is the constriction used in the calculations of the conductance.
}
\end{figure}

The gate potential $U_{\rm QPC}(x,y)$ (see Fig.\ \ref{fig_contour}) defines a saddle-shaped constriction of length $2\ell$, containing a potential barrier of height $V_{\rm QPC}>0$, 
\begin{subequations}
\label{UQPC}
\begin{align}
&U_{\rm QPC}=\begin{cases}
\max\bigl[0,V_{\rm QPC}+U_{\rm saddle}(x,y)\bigr]&
\text{for $|x|>\ell$,}\\
V_{\rm QPC}+ \tfrac{1}{2} m_{\rm eff} \omega_y^2 y^2&
\text{for $|x|<\ell$,}
\end{cases}\label{UQPCa}\\
&U_{\rm saddle}=-\tfrac{1}{2} m_{\rm eff} \omega_x^2 (|x|-\ell)^2 + \tfrac{1}{2} m_{\rm eff} \omega_y^2 y^2.\label{UQPCb}
\end{align}
\end{subequations}
The center $(0,0)$ of the constriction is placed in the normal region at a distance $d$ from the NS interface at $x=d$. The characteristic width $w$ of the constriction at the Fermi energy $E_{F}> V_{\rm QPC}$ is defined by
\begin{equation}
w=\sqrt{\frac{2(E_{F}-V_{\rm QPC})}{m_{\rm eff}\omega_{y}^{2}}}.\label{wdef}
\end{equation}
(This is the separation of classical turning points in the absence of Rashba and Zeeman effects.)

All material parameters have the same value throughout the wire, except the superconducting order parameter $\Delta$, which is set to zero for $x<d$ and $x>L+d$. The length $L$ of the superconducting region if chosen long enough that quasiparticle transmission through it can be neglected (transmission probability $<10^{-7}$). 

Using the algorithm of Ref.\ \cite{Wim09} we calculate the reflection matrix $r$ of the NS junction, which is unitary in the absence of transmission through the superconductor. Andreev reflection is described by the $N\times N$ subblock $r_{he}$,
\begin{equation}
r=\begin{pmatrix}
r_{ee}&r_{eh}\\
r_{he}&r_{hh}
\end{pmatrix}.\label{rblock}
\end{equation}
The Andreev reflection eigenvalues $R_{n}$ ($n=1,2,\ldots N$) are the eigenvalues of the Hermitian matrix product $r_{he}^{\vphantom{\dagger}}r_{he}^{\dagger}$. They are evaluated at the Fermi level for the conductance \eqref{BTK} or at an energy $eV$ above the Fermi level for the differential conductance \eqref{BTKfiniteV}.

\section{B\'{e}ri degeneracy}
\label{app_Beri}

We give a self-contained derivation of the degeneracy of the Andreev reflection eigenvalues discovered by B\'{e}ri \cite{Ber09b,note}. 

The Hamiltonian \eqref{HBdG} satisfies the particle-hole symmetry relation
\begin{equation}
\begin{pmatrix}
0&1\\
1&0
\end{pmatrix}{\cal H}^{\ast}\begin{pmatrix}
0&1\\
1&0
\end{pmatrix}=-{\cal H}.\label{Hehsym}
\end{equation}
For the reflection matrix $r(\varepsilon)$ at energy $\varepsilon $ (relative to the Fermi level) this implies
\begin{equation}
\begin{pmatrix}
0&1\\
1&0
\end{pmatrix}r(\varepsilon)^{\ast}\begin{pmatrix}
0&1\\
1&0
\end{pmatrix}=r(-\varepsilon).\label{rehsym}
\end{equation}
At the Fermi level ($\varepsilon =0$) the electron and hole subblocks in Eq.\ \eqref{rblock} are therefore related by
\begin{equation}
r_{hh}^{\vphantom{\ast}}=r_{ee}^{\ast},\;\;r_{eh}^{\vphantom{\ast}}=r_{he}^{\ast}.\label{ehsymmetry}
\end{equation}

Let us first assume that all $R_{n}$'s are nonzero, so that the matrix $r_{he}$ is invertible. Unitarity $r^{\dagger}r=\openone$ requires that $r_{eh}^{\dagger}r_{ee}^{\vphantom{\dagger}}+r_{hh}^{\dagger}r_{he}^{\vphantom{\dagger}}=0$, hence at the Fermi level
\begin{equation}
A\equiv r_{ee}^{\vphantom{-1}}r_{he}^{-1} =-A^{\rm T}\label{AT}
\end{equation}
is an antisymmetric matrix. (The superscript T denotes the transpose.) The Hermitian matrix product
\begin{equation}
A^{\dagger}A=(r_{he}^{\vphantom{\dagger}}r_{he}^{\dagger})^{-1}-1\label{AA}
\end{equation}
has eigenvalues $a_{n}=1/R_{n}-1$, $n=1,2,\ldots N$.

Let $\Psi$ be an eigenvector of $A^{\dagger}A$ with (real, non-negative) eigenvalue $a$, so $A^{\dagger}A\Psi=a\Psi$. Then $\Psi'=(A\Psi)^{\ast}$ satisfies $A^{\dagger}A\Psi'=-A^{\ast}AA^{\ast}\Psi^{\ast}=A^{\ast}(A^{\dagger}A\Psi)^{\ast}=(aA\Psi)^{\ast}=a\Psi'$. The eigenvalue $a$ is therefore twofold degenerate, unless $\Psi'$ and $\Psi$ are linearly dependent.

If $\Psi'=\lambda\Psi$ for some $\lambda$, then $a\Psi=A^{\dagger}A\Psi=-A^{\ast}(\lambda\Psi)^{\ast}=-|\lambda|^{2}\Psi$, hence $a=0$. So any eigenvalue $1/R_{n}-1\neq 0$ of $A^{\dagger}A$ is twofold degenerate, which implies that the Andreev reflection eigenvalues $R_{n}\neq 0,1$ are twofold degenerate.

To extend the proof to the case that some $R_{n}$'s are zero, we regularize the inverse and consider the matrix
\begin{equation}
A_{\epsilon}=X^{\rm T}r_{he}^{\rm T}r_{ee}^{\vphantom{\rm T}}X,\;\;X=r_{he}^{\dagger}(r_{he}^{\vphantom{\dagger}}r_{he}^{\dagger}+\epsilon)^{-1},\label{Aregularized}
\end{equation}
with $\epsilon$ a positive infinitesimal. This matrix reduces to the one defined in Eq.\ \eqref{AT} if $r_{he}$ is invertible and is well defined even if it is not. Since $A_{\epsilon}$ remains antisymmetric, we can follow the same steps to conclude that the nonzero eigenvalues of $A_{\epsilon}^{\dagger}A_{\epsilon}^{\vphantom{\dagger}}$ are twofold degenerate. Evaluation of this matrix product using the identity
\begin{equation}
r_{ee}^{\dagger}(r_{he}^{\vphantom{\dagger}}r_{he}^{\dagger})^{\rm T}=(r_{he}^{\dagger}r_{he}^{\vphantom{\dagger}})r_{ee}^{\dagger}\label{reerhe}
\end{equation}
gives the expression
\begin{equation}
A_{\epsilon}^{\dagger}A_{\epsilon}^{\vphantom{\dagger}}=(1-r_{he}^{\vphantom{\dagger}}r_{he}^{\dagger})(r_{he}^{\vphantom{\dagger}}r_{he}^{\dagger})^{3}(r_{he}^{\vphantom{\dagger}}r_{he}^{\dagger}+\epsilon)^{-4},\end{equation}
which has eigenvalues $a_{\epsilon,n}=(1-R_{n})R_{n}^{3}(R_{n}+\epsilon)^{-4}$. These are either zero or twofold degenerate, hence we conclude that the $R_{n}$'s are either equal to $0$ or $1$ or twofold degenerate.

Notice that this B\'{e}ri degeneracy is distinct from the familiar Kramers degeneracy (although the proof goes along similar lines \cite{Bar08}).  Kramers degeneracy is a consequence of an anti-unitary symmetry which squares to $-1$. The particle-hole symmetry operation
\begin{equation}
{\cal O}_{\rm ph}=\begin{pmatrix}
0&1\\
1&0
\end{pmatrix}\times \text{complex conjugation}\label{Odef}
\end{equation}
is anti-unitary, but squares to $+1$.

In the absence of time-reversal and spin-rotation symmetry, only the B\'{e}ri degeneracy of the Andreev reflection eigenvalues is operative. This is the case for the model Hamiltonian \eqref{HBdG} considered here (with time-reversal symmetry broken by the Zeeman effect and spin-rotation symmetry broken by the Rashba effect). As worked out in Ref.\ \cite{Bee10}, if one or both of these symmetries are present, then all $R_{n}$'s are twofold degenerate --- including those equal to $0$ or $1$. The Kramers degeneracy then comes in the place of the B\'{e}ri degeneracy, it is not an additional degeneracy.


\begin{thebibliography}{99}
\bibitem{note} The derivation in App.\ \ref{app_Beri} closes a loophole in the proof which we published in New J. Phys. \textbf{13}, 053016 (2011). There we showed that the nonzero product $R_{n}(1-R_{n})$ is twofold degenerate --- leaving open the possibility that the $R_{n}$'s themselves are nondegenerate. We are indebted to C.-Y. Hou for spotting this flaw.
\end{thebibliography}
\end{document}